\documentclass[10pt]{article}
\begin{document}
\title{{\bf Hawking radiation in Reissner-Nordstr\"{o}m blackhole
with a global monopole via Covariant anomalies and Effective action
}}
\author{
{Sunandan Gangopadhyay$^{}$\thanks{sunandan@bose.res.in}}\\
S.~N.~Bose National Centre for Basic Sciences,\\JD Block, 
Sector III, Salt Lake, Kolkata-700098, India\\[0.3cm]
}
\date{}

\maketitle

\begin{abstract}
\noindent 
We adopt the covariant anomaly cancellation method as well
as the effective action approach to obtain the Hawking 
radiation from the Reissner-Nordstr\"{o}m blackhole 
with a global monopole falling in the class of the most 
general spherically symmetric charged blackhole
$(\sqrt{-g}\neq1)$, using only covariant boundary condition at the event
horizon. 
\\[0.3cm]
\\[0.3cm]

\end{abstract}

\noindent {\it{Introduction :}}

\noindent Hawking radiation is an important and prominent quantum effect 
arising from the quantization of matter fields 
in a background spacetime with an event
horizon. The radiation is found to have a spectrum 
with Planck distribution giving
the blackholes one of its thermodynamic properties.
Apart from the original derivation by Hawking \cite{Hawking:sw, Hawking:rv}, 
there is a tunneling picture \cite{Parikh:1999mf, tunneling}
based on pair creations of particles and antiparticles near the horizon
and calculates WKB amplitudes for classically forbidden paths. 
A common feature in these derivations is the universality of the radiation: 
i.e. Hawking radiation is determined universally by the horizon properties
(if we neglect the grey body factor induced by the effect of scattering
outside the horizon). 


\noindent Recently, Robinson and Wilczek (\cite{rw})
proposed an interesting approach to derive Hawking
radiation from a Schwarzschild-type 
black hole through gravitational anomaly.
The method was soon extended to the case of 
charged blackholes \cite{iso}. Further applications of this
approach may be found in \cite{Muratasoda1}-\cite{das1}. 
The basic idea in \cite{rw,iso} is
that the effective theory near the horizon becomes 
two-dimensional and chiral. This chiral theory is anomalous. 
Using the form for two-dimensional 
consistent gauge/gravitational anomaly, Hawking fluxes
are obtained. However the boundary condition necessary to fix the
parameters are obtained from a vanishing of covariant current and 
energy-momentum tensor at the horizon. Soon after, 
the analysis of \cite{rw,iso} was reformulated in \cite{rb}, \cite{sgsk}
in terms of covariant expressions only. 
The generalization of this approach to higher
spin field has been done in \cite{isorecent}.    

\noindent An alternative derivation of 
Hawking flux based  on effective action
using only covariant anomaly has been discussed 
in \cite{rb2}, \cite{sg}, \cite{shail}.
This approach is particularly useful since only the exploitation
of known structure of effective action near the horizon is sufficient 
to determine the Hawking flux. An important ingredient in this method
is once again to realize that the effective theory near the event 
horizon becomes two-dimensional and chiral. Another important
aspect in this approach is the imposition of covariant boundary
conditions only at the horizon.

\noindent In this paper, we first adopt the covariant anomaly cancellation
approach (\cite{rb}) to discuss Hawking radiation 
from Reissner-Nordstr\"{o}m blackhole with a global
monopole \cite{vil} which is an example of the 
most general spherically symmetric charged 
blackhole spacetime ($\sqrt{-g}\neq1$).
\noindent
Finally we adopt the effective action approach \cite{rb2} to
reproduce the same result.
However, as in \cite{sg}, we shall once again ignore
effects to the Hawking flux due to scatterings by the gravitational
potential, for example the greybody factor \cite{kim}.\\

\newpage
\noindent {\it{Hawking radiation from Reissner-Nordstr\"{o}m blackhole 
with a global monopole :}}\\

\noindent The metric of a general non-extremal 
Reissner-Nordstr\"{o}m blackhole with a global monopole $O(3)$ is given by
\cite{vil}
\begin{equation}
ds^{2}_{string} = p(r)dt^{2} - \frac{1}{h(r)}dr^{2} - r^{2}d\Omega^2
\label{2} 
\end{equation}
where,
\begin{eqnarray}
A = \frac{q}{r}dt\quad,\quad 
p(r)= h(r) = 1 -\eta^2 -\frac{2m}{r} +\frac{q^2}{r^2} 
\label{2a}
\end{eqnarray}
with $m$ being the mass parameter of the blackhole and 
$\eta$ is related to the symmetry breaking
scale when the global monopole is formed during 
the early universe soon after the Big-Bang \cite{JP}.
The event horizon for the above blackhole is situated at   
\begin{eqnarray}
r_{H}&=&(1 -\eta^2)^{-1}\big[m +\sqrt{m^2 -(1 -\eta^2)q^2}\big]~.
\label{hor}
\end{eqnarray}
Now it has been argued in \cite{peng} that since 
the metric (\ref{2}) is no longer asymptotically flat, 
so the well known formula\footnote{Note that a spherically 
symmetric asymptotically bounded space-time metric 
without any loss of generality, can
be cast in the form
$ds^2 = g_{tt}dt^2 +g_{rr}dr^2 +r^2 d\Omega^2$.}
\begin{eqnarray}
\kappa=\frac{1}{2}\sqrt{\frac{-g^{rr}}{g_{tt}}}
\big(g_{tt,~r}\big)\big|_{r = r_H}\quad;\quad g_{tt}=p(r),\quad g^{rr}=-h(r) 
\label{oldsg}
\end{eqnarray}
for computing the surface gravity for a 
general spherically symmetric asymptotically flat metric
becomes problematic to be applied 
in the case described by the metric (\ref{2})
as it does not correspond to the normalized time-like
Killing vector. The correct surface gravity 
of the metric (\ref{2}) is 
\begin{eqnarray}
\kappa=\frac{1}{2\sqrt{1 -\eta^2}}p'(r_H) 
\label{newsg}
\end{eqnarray}
since it corresponds to the normalized time-like Killing vector 
\begin{eqnarray}
l_{(t)}^{\mu} = (1 -\eta^2)^{-1/2}(\partial_t)^{\mu}~.
\label{killing}
\end{eqnarray}
It is for this  reason that the anomaly 
cancellation method as well as the effective action approach 
cannot be immediately used to obtain the
consistent formula of the Hawking temperature 
for the metric (\ref{2}).
Nevertheless, we can do the same analysis in another different way. 
By rescaling $t\to\sqrt{1 -\eta^2}~t$, we can 
rewrite the metric (\ref{2}) as
\begin{eqnarray}
ds^2 &=& f(r)dt^2 -h(r)^{-1}dr^2 +r^2d\Omega^2 \nonumber \\
f(r) &=&(1 -\eta^2)h(r) \, , \qquad 
h(r) = 1 -\eta^2 -\frac{2m}{r} -\frac{q^2}{r^2} 
\label{newmetric}
\end{eqnarray}
and immediately derive the expected 
result for the Hawking temperature 
$T = f'(r_H)/(4\pi\sqrt{1 -\eta^2})$.
Hence, we shall apply the anomaly cancellation method and the effective
action approach to the above form of the metric (\ref{newmetric}).
The important point to note is that the determinant of the above metric
$\sqrt{-g}\neq 1$.\\

\noindent {\it{Anomaly cancellation approach :}}\\

\noindent With the aid of dimensional reduction 
procedure one can effectively describe
a theory with a metric given by the ``$r-t$" sector of the full 
spacetime metric (\ref{newmetric}) near the horizon.

\noindent Now we divide the spacetime into two regions and discuss
the gauge/gravitational anomalies separately.\\

\noindent {\it{Gauge anomaly}}\\

\noindent Since the spacetime
has been divided into two regions, we divide the 
current $J^{\mu}$ into two parts. 
The current outside the horizon denoted by $J^{\mu}_{(o)}$ 
is anomaly free and hence satisfies the conservation law
\begin{equation}
\nabla_{\mu}J^{\mu}_{(o)} = 0~. \label{2.4}  
\end{equation} 
Near the horizon there are only outgoing (right-handed) fields
and the current becomes covariantly anomalous and satisfies \cite{iso} 
\begin{eqnarray}
\nabla_{\mu}J^{\mu}_{(H)}&=& -\frac{e^2}{4\pi}
\bar\epsilon^{\rho\sigma}F_{\rho\sigma}= 
\frac{e^2}{2\pi\sqrt{-g}}\partial_{r}A_{t}
\label{2.3}  
\end{eqnarray}  
where, $\bar\epsilon^{\mu\nu}=
\epsilon^{\mu\nu}/\sqrt{-g}$
and $\bar\epsilon_{\mu\nu}=
\sqrt{-g}\epsilon_{\mu\nu}$ are two
dimensional antisymmetric tensors 
for the upper and lower cases
with $\epsilon^{tr}=\epsilon_{rt}=1$.

\noindent Now outside the horizon, 
the conservation equation (\ref{2.4})
yields the differential equation
\begin{equation}
\partial_{r}(\sqrt{-g}J^
{r}_{(o)}) = 0 \label{2.4aa}
\end{equation}
whereas in the region near the horizon, 
the anomaly equation (\ref{2.3})
leads to the following differential equation
\begin{equation}
\partial_{r}\left(\sqrt{-g}J^{r}_{(H)}\right) 
= \frac{e^2}{2\pi}\partial_{r}A_{t}~.
\label{2.4bb}
\end{equation}      
Solving (\ref{2.4aa}) and (\ref{2.4bb}) 
in the region outside and near
the horizon, we get
\begin{eqnarray}
J^{r}_{(o)}(r) &=& \frac{c_{o}}{\sqrt{-g}}\label{2.5}\\
J^{r}_{(H)}(r) &=& \frac{1}{\sqrt{-g}}
\left(c_{H}+\frac{e^2}{2\pi}\int_{r_H}^{r}
\partial_{r}A_{t}(r)\right)\nonumber\\
&=&\frac{1}{\sqrt{-g}}(c_{H} +\frac{e^2}{2\pi}
\left[A_{t}(r) - A_{t}(r_{H})\right])
\label{2.6}
\end{eqnarray}
where, $c_{o}$ and $c_{H}$ are integration constants.
Now as in \cite{iso}, writing $J^{r}(r)$ as 
\begin{equation}
J^{r}(r) = J^{r}_{(o)}(r)\Theta(r-r_{H} -\epsilon) + J^{r}_{(H)}(r) H(r)
\label{2.7}
\end{equation}
where, $ H(r) = 1 - \theta(r-r_{H}-\epsilon)$, we find 
\begin{eqnarray}
\nabla_{\mu}J^{\mu}&=& 
\partial_{r}J^{r}(r)+\partial_{r}(\ln\sqrt{-g}){J^{r}}(r)
\nonumber\\
&=&\frac{1}{\sqrt{-g}}\partial_{r}(\sqrt{-g}{J^{r}}(r))\nonumber\\
&=&\frac{1}{\sqrt{-g}}
\left[\left(\sqrt{-g}(J^{r}_{(o)}(r) - J^{r}_{(H)}(r))
+\frac{e^2}{2\pi}A_{t}(r)\right)\delta(r-r_{+}-\epsilon) +
\partial_{r}\left(\frac{e^2}{2\pi}A_{t}(r) H(r)\right)\right].
\label{2.8}
\end{eqnarray}
The term in the total derivative is cancelled 
by quantum effects of classically irrelevent ingoing modes. 
Hence the vanishing of the Ward identity under gauge transformation 
implies that the coefficient of the delta function is zero, leading
to the condition
\begin{equation}
J^{r}_{(o)}(r) - J^{r}_{(H)}(r))+
\frac{e^2}{2\pi\sqrt{-g}}A_{t}(r)=0~. 
\label{2.90aa}
\end{equation} 
Substituting (\ref{2.5}) and (\ref{2.6}) in the above equation, we get 
\begin{equation}
c_{o} = c_{H} - \frac{e^2}{2\pi}A_{t}(r_{H})~. 
\label{2.9}
\end{equation}                 
The coefficient $c_{H}$ vanishes by requiring that 
the covariant current $J^{r}_{(H)}(r)$ 
vanishes at the horizon. Hence the charge flux
corresponding to $J^{r}(r)$ is given by  
\begin{equation}
c_{o}=\sqrt{-g}J^{r}_{(o)}(r)=-\frac{e^2}{2\pi}A_{t}(r_{H}) 
=-\frac{e^{2}q}{2\pi r_{H}}~.
\label{2.10}
\end{equation}
This is precisely the charge flux obtained in (\cite{peng})
using Robinson-Wilczek method of cancellation of consistent gauge
anomaly. \\ 

\newpage
\noindent {\it{Gravitational anomaly}}\\

\noindent In this case also, since the theory is free from anomaly 
in the region outside the horizon, hence we have the 
energy-momentum tensor satisfying the conservation law
\begin{equation}
\nabla_{\mu}T^{\mu}_{(o)\nu} = F_{\mu\nu}J^{\mu}_{(o)}~.\label{5}
\end{equation}
However, the omission of the ingoing modes in the 
region $r\in[r_{+}, \infty]$ near the horizon, leads to
an anomaly in the energy-momentum tensor there. As we have
mentioned earlier, in this paper we shall focus only on the
covariant form of $d=2$ gravitational anomaly given by (\cite{rw, iso}):
\begin{equation}
\nabla_{\mu}T^{\mu}_{(H)\nu} = F_{\mu\nu}J^{\mu}_{(H)}+\frac{1}{96\pi}
\bar\epsilon_{\nu\mu}\partial^{\mu}R 
= F_{\mu\nu}J^{\mu}_{(H)}+\mathcal A_{\nu}~.
\label{cov}
\end{equation}
It is easy to check that for the metric (\ref{2}), 
the two dimensional Ricci scalar $R$ is given by 
\begin{eqnarray}
R=\frac{h~f^{''}}{f}+\frac{f^{'}h^{'}}{2f}
-\frac{f^{'2}h}{2f^2}
\label{ricci}
\end{eqnarray}
and the anomaly is purely timelike with
\begin{eqnarray}
\mathcal A_{r} &=& 0\nonumber\\
\mathcal A_{t} &=& \frac{1}{\sqrt{-g}}\partial_{r}N^{r}_{t}
\label{8}
\end{eqnarray}
where,
\begin{equation}
N^{r}_{t} = \frac{1}{96\pi}\left(hf'' + 
\frac{f'h'}{2} - \frac{f'^{2}h}{f}\right).
\label{9}
\end{equation}
We now solve the above equations (\ref{5}, \ref{cov}) 
for the $\nu=t$ component.
In the region outside the horizon, the conservation equation (\ref{5})
yields the differential equation
\begin{equation}
\partial_{r}(\sqrt{-g}T^{r}_{(o)t})=\sqrt{-g}F_{rt}J^{r}_{(o)}(r)
=c_{o}\partial_{r}A_{t}  
\label{5a}
\end{equation}
where we have used $F_{rt}=\partial_{r}A_{t}$ and (\ref{2.5}). 
Integrating the above equation leads to
\begin{equation}
T^{r}_{(o)t}(r) = \frac{1}{\sqrt{-g}}(a_{o}+c_{o}A_{t}(r))
\label{6}
\end{equation}
where, $a_{o}$ is an integration constant.
In the region near the horizon, the anomaly equation (\ref{cov})
leads to the following differential equation
\begin{eqnarray}
\partial_{r}\left(\sqrt{-g}T^{r}_{(H)t}\right)&=& 
\sqrt{-g}F_{rt}J^{r}_{(H)}(r)
+\partial_{r}N^{r}_{t}(r)\nonumber\\
&=&(c_{H} +\frac{e^2}{2\pi}
\left[A_{t}(r) - A_{t}(r_{H})\right])\partial_{r}A_{t}(r)
+\partial_{r}N^{r}_{t}(r)\nonumber\\
&=&\partial_{r}\left(\frac{e^2}{2\pi}\left[\frac{1}{2}A^2_{t}(r)-
A_{t}(r_{H})A_{t}(r)\right]+N^{r}_{t}(r)\right)
\label{900}
\end{eqnarray}      
where we have used (\ref{2.6}) in the 
second line and set $c_{H}=0$ in the last line of the above equation.
Integration of the above equation leads to
\begin{eqnarray}
T^{r}_{(H)t}(r)&=&\frac{1}{\sqrt{-g}}\left(b_{H}+\int_{r_H}^{r}
\partial_{r}\left(\frac{e^2}{2\pi}
\left[\frac{1}{2}A^{2}_{t}(r) - A_{t}(r_{H})A_{t}(r)\right]
+N^{r}_{t}(r)\right)
\right)
\nonumber\\
&=& \frac{1}{\sqrt{-g}}
\left(b_{H} +\frac{e^2}{4\pi}[A^2_{t}(r)+A^2_{t}(r_H)]
-\frac{e^2}{2\pi}A_{t}(r_{H})A_{t}(r)+ 
N^{r}_{t}(r) - N^{r}_{t}(r_{H})\right)
\label{10}
\end{eqnarray}
where, $b_{H}$ is an integration constant. 

\noindent Writing the energy-momentum tensor as a 
sum of two contributions \cite{iso}
\begin{equation}
{T^{r}}_{t}(r) = T^{r}_{(o)t}(r)\theta(r-r_{H}-\epsilon) 
+ T^{r}_{(H)t}(r)H(r)
\label{11}
\end{equation}
we find 
\begin{eqnarray}
\nabla_{\mu}{T^{\mu}}_{t}
&=&\partial_{r}{T^{r}}_{t}(r)+\partial_{r}(\ln\sqrt{-g}){T^{r}}_{t}(r)
\nonumber\\
&=&\frac{1}{\sqrt{-g}}\partial_{r}(\sqrt{-g}{T^{r}}_{t}(r))\nonumber\\
&=&\frac{1}{\sqrt{-g}}\left[-\frac{e^2}{2\pi}A_{t}(r_H)\partial_{r}A_{t}(r)
+\left(\sqrt{-g}(T^{r}_{(o)t}(r) - T^{r}_{(H)t}(r))
+\frac{e^2}{4\pi}A^{2}_{t}(r)
+N^{r}_{t}(r)\right)\delta(r-r_{+}-\epsilon)\right.\nonumber\\ 
&&\left.+\partial_{r}\left([\frac{e^2}{4\pi}A^{2}_{t}(r)+N^{r}_{t}(r)] H(r)
\right)\right]
\label{12} 
\end{eqnarray}
where we have substituted the value of $c_0$ from (\ref{2.10})
in the last line.

\noindent Now the first term in the above equation 
is a classical effect coming from the Lorentz force. 
The term in the total derivative is once again 
cancelled by quantum effects of classically irrelevant ingoing modes. 
The quantum effect to cancel this
term is the Wess-Zumino term induced by the ingoing modes near the horizon. 
Hence the vanishing of the Ward identity under diffeomorphism transformation
implies that the coefficient of the delta function in the above 
equation vanishes
\begin{equation}
T^{r}_{(o)t} - T^{r}_{(H)t} + 
\frac{1}{\sqrt{-g}}(\frac{e^2}{4\pi}A^{2}_{t}(r)+N^{r}_{t}(r)) = 0~.
\label{14}
\end{equation}
Substituting (\ref{6}) and (\ref{10}) in the above equation, 
we get
\begin{equation}
a_{o} = b_{H}+\frac{e^2}{4\pi}A^{2}_{t}(r_H) - N^{r}_{t}(r_{H})~.
\label{15}
\end{equation}
The integration constant $b_{H}$ can be fixed by imposing 
that the covariant energy-momentum tensor vanishes at the horizon. 
From (\ref{10}), this gives $b_{H}=0$.
Hence the total flux of the energy-momentum tensor is given by
\begin{eqnarray}
a_{o}&=&\frac{e^2}{4\pi}A^{2}_{t}(r_H)-N^{r}_{t}(r_{H})\nonumber\\
&=&\frac{e^{2}q^2}{4\pi r^2_H}+\frac{1}{192\pi} f'(r_{H})h'(r_{H})\nonumber\\
&=&\frac{e^{2}q^2}{4\pi r^2_H}+\frac{1}{192\pi}\frac{f'^2(r_{H})}{(1-\eta^2)}~.
\label{flux}
\end{eqnarray}\\
This is precisely the Hawking flux obtained in (\cite{peng})
using Robinson-Wilczek method of cancellation of consistent
anomaly. \\

\noindent {\it{Effective action approach :}}\\

\noindent As we have already mentioned earlier, 
with the aid of dimensional reduction technique,
the effective field theory  near the 
horizon becomes a two dimensional chiral theory 
with a metric given by the ``$r- t$" sector of the full spacetime metric
(\ref{newmetric}) near the horizon.

\noindent We now adopt the methodology in \cite{rb2}.
For a two dimensional theory the expressions 
for the anomalous (chiral) and normal effective actions 
are known \cite{Leut}. 
We shall use only the anomalous form 
of the effective action for 
deriving the charge and the energy flux.
The current and the energy-momentum tensor in the region near the horizon
is  computed by taking appropriate functional derivative of the chiral
effective action. Next, the parameters appearing
in the solution is fixed by imposing 
the vanishing of covariant current and energy-momentum
tensor at the horizon. Once these are fixed, 
the charge and the energy flux are obtained by taking the asymptotic 
$(r\rightarrow {\infty})$ limit of the chiral current and energy-momentum 
tensors. We also use the expression for the normal effective
action to establish a connection between the chiral and the
normal current and energy-momentum tensors.

\noindent With the above methodology in mind, 
we write down the anomalous (chiral) effective action (describing the theory 
near the horizon) \cite{Leut} 
\begin{equation}
\Gamma_{(H)}= -\frac{1}{3} z(\omega)+z(A) 
\label{effaction}
\end{equation}
where $A_{\mu}$ and $\omega_{\mu}$ are the gauge field and
the spin connection and
\begin{equation}
z(v) = \frac{1}{4\pi}\int d^2x~ d^2y~ \epsilon^{\mu\nu}
\partial_\mu v_\nu(x) \Delta_{g}^{-1}(x, y)
\partial_\rho[(\epsilon^{\rho\sigma} + \sqrt{-g}g^{\rho\sigma})v_\sigma(y)]
\label{effaction1}
\end{equation}
where $\Delta_{g} = \nabla^{\mu}\nabla_{\mu}$ is the laplacian in this 
background.

\noindent The energy-momentum 
tensor is computed from a variation of this effective action.
To get their covariant forms in which we are interested, 
one needs to add appropriate local polynomials \cite{Leut}. 
Here we quote the final result for the chiral covariant 
energy-momentum tensor and the chiral covariant current \cite{Leut}: 
\begin{eqnarray}
{T^{\mu}}_{\nu} = \frac{e^2}{4\pi}D^{\mu}B D_{\nu}B 
+\frac{1}{4\pi}\left(\frac{1}{48}D^{\mu}G D_{\nu}G 
-\frac{1}{24} D^{\mu} D_{\nu}G + \frac{1}{24}\delta^{\mu}_{\nu}R\right)
\label{eff3} 
\end{eqnarray}
\begin{eqnarray}
J^{\mu}= -\frac{e^2}{2\pi}D^{\mu}B \label{eff4} 
\end{eqnarray}
where $D_{\mu}$ is the chiral covariant derivative
\begin{equation}
D_{\mu} = \nabla_{\mu} - \bar{\epsilon}_{\mu\nu}\nabla^{\nu}
= -\bar{\epsilon}_{\mu\nu}D^{\nu}~. \label{eff5}
\end{equation} 
Also $B(x)$ and $G(x)$ are given by
\begin{eqnarray}
B(x) &=& \int d^2y  \ \sqrt{-g}\Delta^{-1}_{g}(x,y)\bar{\epsilon}^{\mu\nu}
\partial_{\mu}A_{\nu}(y)
\label{eff7}
\end{eqnarray}
\begin{eqnarray}
G(x) &=& \int d^2y  \ \Delta^{-1}_{g}(x,y)\sqrt{-g}~R(y)\label{eff6}
\end{eqnarray}
and satisfy
\begin{equation}
\nabla^{\mu}\nabla_{\mu}B =-\partial_{r}A_{t}(r)
\quad,\quad \nabla^{\mu}\nabla_{\mu}G = R ~.
\label{e1}
\end{equation}
The solutions for $B$ and $G$ read 
\begin{equation}
B = B_o(r) - at + b \ ; \  \partial_{r}B_o = \frac{1}{\sqrt{fh}}
\left(A_{t}(r)+c\right)
\label{sol1}
\end{equation}\begin{equation}
G = G_o(r) - 4 pt + q \ ; \  \partial_{r}G_o = - \frac{1}{\sqrt{fh}}
\left(\sqrt{\frac{h}{f}}f'+z\right)
\label{sol2}
\end{equation}
where $a,b,c, p, q, z$ are constants of integration. 

\noindent By taking the covariant divergence 
of (\ref{eff3}) and (\ref{eff4}), 
we get the anomalous Ward identities
(\ref{cov}) and (\ref{2.3}).
 
\noindent In the region away from the horizon, the effective
theory is given by the standard effective action
$\Gamma$ of a conformal field
with a central charge $c=1$ 
in this blackhole background \cite{Leut} and reads:
\begin{eqnarray}
\Gamma =  \frac{1}{96\pi}\int d^2x d^2y \ \sqrt{-g}~ 
R(x)\frac{1}{\Delta_{g}}(x,y)\sqrt{-g}~R(y)
+\frac{e^2}{2\pi}\int d^2x d^2y \epsilon^{\mu\nu}
\partial_{\mu}A_{\nu}(x)\frac{1}{\Delta_{g}}(x,y)
\epsilon^{\rho\sigma}\partial_{\rho}A_{\sigma}(y)~. 
\label{normaleff}
\end{eqnarray}
The covariant energy-momentum tensor $T_{\mu\nu(o)}$ and the 
covarant gauge current $J^{\mu}_{(o)}$ in the region outside 
the horizon are given by
\begin{eqnarray}
T_{\mu\nu(o)}&=&\frac{2}{\sqrt{-g}} \frac{\delta\Gamma}{\delta g^{\mu\nu}}
\nonumber\\
&=&\frac{1}{48\pi} \left(2g_{\mu\nu}R - 2\nabla_{\mu}\nabla_{\nu}G
+\nabla_{\mu}G\nabla_{\nu}G-\frac{1}{2}g_{\mu\nu}
\nabla^{\rho}G\nabla_{\rho}G\right)\nonumber\\
&&+\frac{e^2}{\pi}\left(\nabla_{\mu}B\nabla_{\nu}B
-\frac{1}{2}g_{\mu\nu}
\nabla^{\rho}B\nabla_{\rho}B\right)
\label{normalten}
\end{eqnarray}
\begin{eqnarray}
J^{\mu}_{(o)}&=&\frac{\delta\Gamma}{\delta A^{\mu}}
=\frac{e^2}{\pi}\bar{\epsilon}^{\mu\nu}\partial_{\nu}B
\label{norten}
\end{eqnarray}
and satisfy the normal Ward identities (\ref{5}) and (\ref{2.4}).\\

\noindent{\it Charge and Energy Flux:}\\

\noindent In this section we calculate the charge and the energy 
flux by using the expressions for 
the anomalous covariant gauge current (\ref{eff4}) and 
anomalous covariant energy-momentum tensor (\ref{eff3}).
We will show that the results are the
same as that  obtained by the covariant anomaly cancellation 
method. 
 
\noindent Using the solution for $B(x)$ (\ref{eff7}) and (\ref{eff5}), 
the $\mu=r$ component of the anomalous (chiral) covariant 
gauge current (\ref{eff4}) becomes
\begin{eqnarray}
J^{r}(r)&=&\frac{e^2}{2\pi}\sqrt{\frac{h}{f}}
[A_{t}(r)+a+c].
\label{bb1}
\end{eqnarray}
Now implementation of the boundary condition 
namely the vanishing of the anomalous (chiral) covariant gauge current
at the horizon, $J^{r}(r_{H}) = 0$, leads to
\begin{equation}
a+c=-A_{t}(r_{H})~.
\label{bb2} 
\end{equation} 
Hence the expression $J^{r}(r)$ reads
\begin{eqnarray}
J^{r}(r)&=&\frac{e^2}{2\pi}\sqrt{\frac{h}{f}}
[A_{t}(r)-A_{t}(r_{H})].
\label{bb3}
\end{eqnarray}
\noindent Now the charge flux is given by the 
asymptotic ($r\rightarrow\infty$) limit of the
anomaly free covariant gauge current (\ref{norten}). Now from 
(\ref{2.3}), we observe that the anomaly vanishes
in this limit. Hence the charge flux is abstracted
by taking the asymptotic limit of the above equation multiplied
by an overall factor of $\sqrt{-g}$.
This yields
\begin{equation}
c_0=(\sqrt{-g}J^{r})(r\rightarrow\infty)
=(\sqrt{\frac{f}{h}}
J^{r})(r\rightarrow\infty)
=-\frac{e^2}{2\pi}A_{t}(r_{H})
=-\frac{e^2 q}{2\pi r_{H}}
\label{bb4}
\end{equation}
which agrees with (\ref{2.10}).

\noindent We now consider the normal (anomaly free) covariant
gauge current (\ref{norten}) to establish its relation with
the chiral (anomalous) covariant gauge current 
(\ref{bb3}). The $\mu=r$ component of $J^{\mu}_{(o)}$
is given by
\begin{eqnarray}
J^{r}_{(o)}(r)&=&\frac{e^2}{\pi}\sqrt{\frac{h}{f}}~a~.
\label{bb5}
\end{eqnarray}
The asymptotic form of the above equation (\ref{bb5}) must agree
with the asymptotic form of (\ref{bb1})\footnote{This is true
since the anomaly in the asymptotic limit ($r\rightarrow\infty$) 
vanishes as can be readily seen from (\ref{2.3}).}. 
This yields (using (\ref{bb2})):
\begin{eqnarray}
a=c=-\frac{1}{2}A_{t}(r_{H})~.
\label{bb10}
\end{eqnarray}
Using the above solutions in (\ref{bb1}) 
and (\ref{bb5}) yields (\ref{bb3}) and
\begin{eqnarray}
\sqrt{-g}J^{r}_{(o)}(r)=
\sqrt{\frac{f}{h}}J^{r}_{(o)}(r) =-\frac{e^2}{2\pi}A_{t}(r_{H})~.
\label{bb7}
\end{eqnarray} 
The above expressions (\ref{bb3}) and (\ref{bb7}) 
yields the equation between the chiral (anomalous) and the normal
energy-momentum tensors (\ref{2.90aa}).

\noindent Now we focus our attention on the gravity sector.
Using the solutions for $B(x)$ (\ref{eff7}) and $G(x)$ (\ref{eff6}), 
the $r-t$ component of the anomalous (chiral) covariant 
energy-momentum tensor (\ref{eff3}) becomes
\begin{eqnarray}
{T^{r}}_t(r)&=&\frac{e^2}{4\pi}\sqrt{\frac{h}{f}}
[A_{t}(r)-A_{t}(r_{H})]^2+\frac{1}{12\pi}\sqrt{\frac{h}{f}}
\left[p - \frac{1}{4}\left(\sqrt{\frac{h}{f}}f' + z\right)\right]^2 
\nonumber\\
&&+\frac{1}{24\pi}\sqrt{\frac{h}{f}}
\left[\sqrt{\frac{h}{f}}f'
\left(p-\frac{1}{4}\left(\sqrt{\frac{h}{f}}f' + z\right)\right) 
+ \frac{1}{4}hf'' - \frac{f'}{8}\left(\frac{h}{f}f'-h'\right)
\right].\label{eflux1}
\end{eqnarray}
Now implementing the boundary condition 
namely the vanishing of the covariant energy-momentum tensor
at the horizon, ${T^{r}}_t(r_{H}) = 0$, leads to
\begin{equation}
p=\frac{1}{4}\left[z \pm \sqrt{f'(r_H)h'(r_H)}\right] \ ; 
\  f'(r_H) \equiv f'(r = r_{H})~. 
\label{efl2} 
\end{equation} 
Using either of the above solutions in (\ref{eflux1}) yields
\begin{eqnarray}
{T^{r}}_t(r) =\frac{e^2}{4\pi}\sqrt{\frac{h}{f}}
[A_{t}(r)-A_{t}(r_{H})]^2+\frac{1}{192\pi}\sqrt{\frac{h}{f}}
\left[f'(r_H)h'(r_H) - 
\frac{2h}{f}f'^{2} + 2hf''+f'h'\right]. 
\label{efl3}
\end{eqnarray} 
The energy flux is now given by the 
asymptotic ($r\rightarrow\infty$) limit of the
anomaly free energy-momentum tensor (\ref{normalten}). Now from 
(\ref{cov}), we observe that the anomaly vanishes
in this limit. Hence the energy flux is abstracted
by taking the asymptotic limit of the above equation multiplied
by an overall factor of $\sqrt{-g}$.
This yields
\begin{eqnarray}
a_0&=&(\sqrt{-g}{T^{r}}_t)(r\rightarrow\infty)
=(\sqrt{\frac{f}{h}}{T^{r}}_t)(r\rightarrow\infty)\nonumber\\
&=&\frac{e^2}{4\pi}A_{t}(r_{H})]^2+\frac{1}{192\pi}f'(r_H)h'(r_H)\nonumber\\
&=&\frac{e^2 q^2}{4\pi r_{H}^2}
+\frac{1}{192\pi}f'(r_H)h'(r_H)
\label{efl4}
\end{eqnarray}
which correctly reproduces the energy flux (\ref{flux}).

\noindent We now consider the normal (anomaly free) energy-momentum
tensor (\ref{normalten}) to establish its relation with
the chiral (anomalous) energy-momentum
tensor (\ref{efl3}). The $r-t$ component of ${T^{\mu}}_{\nu(o)}$
is given by
\begin{eqnarray}
{T^{r}}_{t(o)}(r)&=&\frac{e^2}{\pi}\sqrt{\frac{h}{f}}
a[A_{t}(r)+c]-\frac{1}{12\pi}\sqrt{\frac{h}{f}}zp\nonumber\\
&=&-\frac{e^2}{2\pi}\sqrt{\frac{h}{f}}A_{t}(r_{H})[A_{t}(r)
-\frac{1}{2}A_{t}(r_{H})]-\frac{1}{12\pi}\sqrt{\frac{h}{f}}zp
\label{rtnormal}
\end{eqnarray}
where we have used (\ref{bb10}).
Once again since the anomaly in the asymptotic limit ($r\rightarrow\infty$) 
vanishes as can be readily seen from (\ref{cov}), 
the asymptotic form of the above equation (\ref{rtnormal}) must agree
with the asymptotic form of (\ref{eflux1}).
This yields:
\begin{eqnarray}
p=-\frac{z}{4}~.
\label{asym}
\end{eqnarray}
Solving (\ref{efl2}) and (\ref{asym}) gives two solutions for $p$ and $z$ :
\begin{eqnarray}
p&=&\frac{1}{8}\sqrt{f'(r_H)h'(r_H)}\quad;\quad 
z=-\frac{1}{2}\sqrt{f'(r_H)h'(r_H)}\nonumber\\
p&=&-\frac{1}{8}\sqrt{f'(r_H)h'(r_H)}\quad;\quad 
z=\frac{1}{2}\sqrt{f'(r_H)h'(r_H)}~. 
\label{solutions}
\end{eqnarray}
Using either of the above solutions in (\ref{eflux1}) 
and (\ref{rtnormal}) yields (\ref{efl3}) and
\begin{eqnarray}
\sqrt{-g}{T^{r}}_{t(o)}(r)=\sqrt{\frac{f}{h}}{T^{r}}_{t(o)}(r)
=-\frac{e^2}{2\pi}A_{t}(r_{H})[A_{t}(r)
-\frac{1}{2}A_{t}(r_{H})]+\frac{1}{192\pi}
f'(r_H)h'(r_H)~. 
\label{energy10}
\end{eqnarray} 
The above expressions (\ref{efl3}) and (\ref{energy10}) 
yields the equation between the chiral (anomalous) and the normal
energy-momentum tensors (\ref{14}).\\

\noindent {\it{Discussions :}}

\noindent In this paper, we studied the problem of Hawking radiation 
from Reissner-Nordstr\"{o}m blackhole with a global monopole
using covariant anomaly cancellation technique and effective action approach.
The point to note in the anomaly cancellation method 
is that Hawking radiation plays
the role of cancelling gauge and gravitational
anomalies at the horizon to restore the gauge/diffeomorphism
symmetry at the horizon.
An advantage of this method is that neither
the consistent anomaly nor the counterterm relating the different
(covariant and consistent) currents, 
which were essential ingredients in \cite{iso},
were required. 

\noindent On the other hand, in the effective action technique, 
we only need covariant boundary conditions, the
importance of which was first stressed in \cite{rb2}.
Another important input in the entire procedure is the expression
for the anomalous (chiral) effective action 
(which yields anomalous Ward identity 
having covariant gauge/gravitational anomaly). 
The unknown parameters in the covariant current and energy-momentum tensor 
derived from this anomalous effective action 
were fixed by a boundary condition- namely the vanishing of the 
covariant current and energy-momentum tensor 
at the event horizon of the blackhole.
Finally, the charge and the energy flux were extracted 
by taking the $r\rightarrow\infty$
limit of the chiral covariant current and energy-momentum tensor.
The relation between the chiral and the normal current and 
energy-momentum tensors is also established by requiring that 
both of them match in the 
asymptotic limit which is possible since the gauge/gravitational 
anomaly vanishes in this limit.\\

\section*{Acknowledgements }
The author would like
to thank Prof. R. Banerjee and Shailesh Kulkarni for useful comments.


\end{document}